\documentstyle[prl,aps,epsfig,multicol]{revtex}
\begin{document}
\def\be{\begin{equation}}
\def\ee{\end{equation}}
\def\bearr{\begin{eqnarray}}
\def\eearr{\end{eqnarray}}
\def\la{\langle}
\def\ra{\rangle}
\def\l{\left}
\def\r{\right}
\draft
\title{ Quantized Hydrodynamic Theory of One-Dimensional Hard Core Bosons}
\author{Tarun Kanti Ghosh}
\address
{ The Abdus Salam International Centre for Theoretical Physics, 
34100 Trieste, Italy.}

\date{\today}
\maketitle

\begin{abstract}
We present a quantized hydrodynamic theory and its applications of one-dimensional 
hard-core bosons in a harmonic trap. Quantizing the Hamiltonian of a trapped hard-core 
bosons and diagonalize it in terms of the phase and density fluctuations associated with 
the Bose field operator. As an applications, we calculate discrete energy spectrums, 
dynamic structure factor, momentum transferred by a two-photon Bragg pulse, single-particle 
density matrix, momentum distribution, and two-particle correlation function. 
%We find that there is a crossing between two consecutive Bogoliubov modes at finite momentum due to the 
%strong non-linear interactions which causes the mixing of different Bogoliubov modes. 
The dynamic structure factor has multiple peaks due to the discrete 
nature of the eigen-modes which can be observed by a two-photon Bragg pulse with long 
duration. The coherence length at zero temperature is very small due to the ``fermionization" of 
the system. We also compute the momentum distributions which has oscillatory behavior at large
momentum. The pair distribution function shows that there are many deep valleys at various 
relative separations which implies shell structure due to the Pauli blocking in real space. 
\end{abstract}


\begin{multicols}{2}[]

\section{Introduction}
Recently, there have been renewed interest in the study 
of many-body effects in a one-dimensional(1D) quantum fluid.
There is a realization of quasi-1D degenerate Bose gas in a 
harmonic trap \cite{gor} by freezing the radial motion of atoms
in a cylindrical trap to zero point oscillations. Depending on the 
strength of the two-body interaction, the trapped bosons are in two regimes:
mean-field regime when the particles are weakly interacting which can be described
by the Gross-Pitaevskii equation, and hard-core bosons (commonly known as 
Tonks-Girardeau gas) when the interaction is strong enough which can be described
by the Lieb-Liniger model of $\delta$-interacting 1D bosons \cite{lieb}.
Most of the study concentrate on the phase fluctuating degenerate bosons 
in the mean-field regime \cite{ho,pet,men,griffin,mora,tk}. With the possibility 
of creation of the hard-core bosonic system in a harmonic trap, there is a growing interest  
to study theoretically this strongly interacting regime 
\cite{kolo,olsha1,tosi,caza,girar1,gang,papen,giorgini,olsha,caza1}. 
In this regime, 
the system acquires fermionic properties in a sense that there exists an 
exact mapping between the density of a strongly interacting bosons and the density 
of an ideal Fermi gas. The energy per particle of hard-core bosons coincides with 
the  Fermi energy of an ideal Fermi gas: $E/N = (\pi \hbar n)^2/(6m)$ \cite{girar}. 
This is a manifestation of the Fermi-Bose duality in 1D systems.

This paper is organised as follows. In section II, we write down an effective hydrodynamic
Hamiltonian of a trapped hard-core bosons and diagonalize it in terms of the phase and density
fluctuations. In section III,
we calculate the dynamic structure factor and the time evolution of the momentum transferred due 
to the two-photon Bragg pulse. In section IV, we calculate the single-particle correlation function,
momentum distributions and the two-particle correlation functions. In section V, we present a brief
summary of our work.

\section{quantization of 1D hard-core bosons}
We consider $N$ interacting bosons in a cigar-shaped trap ($ \omega_z \ll \omega_0$)
at very low temperature $( T \sim 0)$.  
If temperature and the interaction energy per particle does not
exceed the transverse level spacing $ \hbar \omega_0 $, then the elongated 
cigar-shaped trapped system becomes effectively one-dimensional.
The two-body interaction in the axial direction is approximated by an
effective two-body potential \cite{olsha}, $ U_{1D} = g_{1} \delta (z) $, where
$ g_{1} = -\frac{2 \hbar^2}{ m a_{1D}}$ is
an effective one-dimensional coupling constant and $ a_{1D} = - \frac{a_0^2}{ a}[1- 1.4603 \frac{a}{\sqrt{2}a_0}]$
is the one-dimensional scattering length, where $ a $ is the scattering length in $ 3D $ system
and $ a_0 = \sqrt{\frac{\hbar}{m \omega_0}} $ is the transverse oscillator length. 
The effective one-dimensional Hamiltonian for trapped atoms is
\be
H = H_0 + \sum_{i=1}^N \frac{m\omega_z^2 z_i^2}{2},
\ee
where
\be
H_0 = -\frac{\hbar^2}{2m} \sum_{i=1}^N \frac{\partial^2}{\partial z_i^2} + g_1 
\sum_{i=1}^N \sum_{j=i+1}^N \delta(z_i-z_j)
\ee
is known as for a 1D $\delta$-interacting Bose gas. This Hamiltonian
can be diagonalized via Bethe ansatz and the equation of state is known 
exactly for all densities and temperatures.
%However, the wave functions which one obtains from the Bethe ansatz
%are usually quite difficult to work with. It is therefore useful to
%examine other ways of studying the hard-core bosons in a harmonic trap.
%With this in mind, we apply quantized hydrodynamic theory to hard-core 
% bosons in a harmonic trap and examine different observables.
The gas parameter is defined as the product of the density ($n$) with 
the effective scattering length $a_{1D}$, i.e $ n |a_{1D}| $.
At zero temperature, the interaction energy per particle is 
$ \epsilon (n) = \frac{1}{3} \frac{\pi^2 \hbar^2}{2m} n^2$ when $ n |a_{1D}| \rightarrow 0 $.
The low-density limit corresponds to the case of infinitely strong interactions.
We are interested to study the Bose systems with infinitely strong interactions at 
zero temperature by using the quantized hydrodynamic approximation.
Therefore, in the dilute regime, the Hamiltonian for the hard-core bosons confined in
a harmonic trap at zero temperature can be written as,
\bearr \label{hamil}
H^{\prime} & = & \int dz \Phi^* (z,t) ( - \frac{\hbar^2}{2m} \frac{\partial^2}{\partial z^2}
+\frac{1}{2} m \omega_z^2 z^2 - \mu ) \Phi (z,t) \nonumber
\\ & + & \frac{g_{1}}{3} \int dz |\Phi (z,t)|^6,
\eearr  
where $g_1 = (\pi^2 \hbar^2)/(2m)$.
The Thomas-Fermi density profile of the trapped hard-core bosons is different
from bosons in the mean-field regime and it is
\be \label{density}
n_0 (z) = \frac{\sqrt{2N}}{\pi a_z} \sqrt{[1 - \frac{z^2}{Z_0^2}]},
\ee
where $Z_0 = \sqrt{2N} a_z$ is the Thomas-Fermi length of the system.
The chemical potential is $ \mu = N \hbar \omega_z$. Also, one can define the Fermi
wave-vector is $k_f = \sqrt{2N}/a_z = 1/l_c$, where $l_c $ is the correlation length
of this system. We extend the quantized hydrodynamic theory developed by Wu and Griffin \cite{griffin,wu}
at $T=0$ for the bosons in the mean-field regime to the hard-core bosons in a harmonic trap. 
It should be mentioned that the hydrodynamic theory of strongly interacting system is valid 
only when $ |z| > l_c $ or $ k l_c < 1 $.
The Bose order parameter can be written as $ \Phi(z,t) = \sqrt{n(z,t)} e^{i\phi(z,t)}$.
The fluctuations in the density and phase about their equilibrium are 
$ \hat n(z,t) = n_0(z) + \delta \hat n(z,t) $ and $ \hat \phi(z,t) = \phi_0 (z) + \delta \hat \phi(z,t) $. 
By keeping up to second-order in the fluctuations, we obtain the effective hydrodynamic Hamiltonian at
zero temperature,
$$
H = H_0 + \frac{1}{2} \int dz \l [m n_0(z) \delta \hat v_z^2(z,t) + g_1 n_0(z) \delta \hat n^2(z,t) \r],
$$
where $ \delta v_z (z,t) = \frac{\hbar}{m} \frac{d}{dz} \delta \phi(z,t) $ is the superfluid velocity.
This Hamiltonian is similar to the 1D Hamiltonian derived by Haldane \cite{haldane}. 
This Hamiltonian is quadratic in the density and the phase fluctuation operators, then
we can diagonalize it using the canonical transformations:
\be
\delta \hat n (z,t) = \sum_{j} \l [ a_j f_j(z) e^{-i \omega_j t} 
\hat \alpha_j + H. c. \r],
\ee
and
\be
\delta \hat \phi (z,t) = \sum_{j} \l [ b_j \psi_j(z) e^{-i \omega_j t } \hat \alpha_j + H.c. \r],
\ee
where $ f_j(z) = \frac{\psi_j(z)}{\sqrt{1-(z/Z_0)^2}}$.
The operators $ \hat \alpha_j $ and $ \hat \alpha_j^{\dag} $ destroy and 
create excitations with energy $ \hbar \omega_j$
and satisfy the commutation relations $[ \hat \alpha_j, \hat \alpha_{j^{\prime}}^{\dag}] = 
\delta_{j,j^{\prime}} $, 
$ [ \hat \alpha_j, \hat \alpha_j] = 0 $ and $ [ \hat \alpha_j^{\dag}, \hat \alpha_j^{\dag}] = 0 $.
Also, $ [\delta \hat n(z,t), \delta \hat \phi (z^{\prime},t)] =  i \delta (z - z^{\prime}) $.
The constant terms are $ a_j = i \sqrt{\frac{\hbar \omega_j}{4 g_1 n_0(0)}}$ and 
$b_j = \sqrt{\frac{g_1 n_0(0)}{\hbar \omega_j}}$.
The equations for the density and phase fluctuations can be obtained by using the Heisenberg equation
of motion:
\be
\frac{ \partial^2 \delta \hat n(z,t)}{\partial t^2} = \frac{\partial}{\partial z} 
\l [\frac{g_1}{m} n_0(z) \frac{\partial}{\partial z} [ n_0(z) \delta \hat n(z,t)] \r],
\ee
and 
\be
\frac {\partial^2 \delta \hat \phi (z,t)}{\partial t^2} = 
\l [\frac{g_1}{m} n_0(z) \frac{\partial}{\partial z} [ n_0(z) \delta \hat \phi (z,t)] \r].
\ee

The equation for the eigenfunctions $ \psi_j (z) $ is
\be \label{gen1}
[(1- \tilde z^2) \frac{\partial^2}{\partial \tilde z^2} - \tilde z \frac{\partial}{\partial \tilde z} 
+ \omega_j^2 ]\psi_j(z) = 0,
\ee
where $ \tilde z = z/Z_0$.
Therefore, Eq. (\ref{gen1}) becomes Chebyshev equation
with the eigenfrequencies given by $ \omega_j = j \omega_z  $ and the
corresponding normalized eigenfunctions are $ \psi_j (z) = \sqrt{\frac{\delta_j}{\pi Z_0}} T_j(z/Z_0) $,
where $ T_j(z/Z_0) $ is the first-order Chebyshev polynomial in $z$ and  $\delta_0 = 1$, otherwise $\delta_j = 2$.
The monopole mode corresponds to the $j=2$ mode which is independent of the interaction strength
\cite{tk1} due to the conformal symmetry \cite{piju} of the Hamiltonian (see Eq. (\ref{hamil})).
Note that the eigenfunctions of the density and phase fluctuations are different in the hard-core bosons
regime, in contrast to the case of bosons in the mean-field regime.

\section{structure factors and Bragg spectroscopy}
{\em Dynamic Structure Factor}:
Consider a low-intensity off-resonant inelastic
light scatters with momentum transfer $\hbar k $ and energy transfer $\hbar \omega $
to the target (hard-core bosons). If the external light couples weakly to the number density
of the target, the differential cross section is proportional to the dynamic structure factor
$ S(k,\omega) $, which is obtained from the Fourier transform of the time-dependent density-density
correlation functions,
\be
S(k,\omega)  = \int dt \int dz e^{i(\omega t - kz)} < \delta \hat n(z,t) \delta \hat n(0,0)>.
\ee
It is the density fluctuation spectrum that can be measured in the two-photon Bragg 
spectroscopy. 
The dynamic structure factor of this system can be written as,
\bearr
S(k,\omega) & = & \sum_{j=1} \frac{\hbar \omega_j}{4 g_1 n_0(0)} |F_j(k)|^2 \nonumber \\ 
& \times & [ N_j \delta(\omega+\omega_j) +(1+ N_j ) \delta(\omega - \omega_j)],
\eearr
where $ N_j = [exp{(\beta \hbar \omega_j)} - 1]^{-1} $ is the thermal Bose-Einstein
function and $ F_j(k) = \int dz e^{-iqz} f_j(z) $ is the spatial Fourier transformation
of $ f_j(z) $.
It can be easily shown that 
\be
|F_j(\tilde k)|^2 =  Z_0 \pi^2 |J_{j}(2N \tilde k)|^2,
\ee
where the dimensionless variable is $ \tilde k = k/k_f$.
We can write the dynamic structure factor at $T=0 $ as
$ S( \tilde k,\omega) =  \nonumber \sum_{j} S_j(\tilde k) \delta(\omega -\omega_j)] $,
where
\be \label{sjq}
S_j(\tilde k) = j |J_j( 2N \tilde k)|^2.
\ee
These functions determine the weight of the light-scattering cross-section in $S(k,\omega)$
of the corresponding collective modes of energy $\omega_j$.
In Fig.1 we plot $ S_j(\tilde k) $ as a function
of the dimensionless wave vector $ \tilde k = k/k_f $ for the excitations $ j = 1,2,3,4,$ and $ 5 $.
Fig.1 shows that for a given $k$, how many modes significantly contribute to $S(k,\omega)$. 
Note that $j=0$ mode do not contribute to $S(k,\omega)$.
It is clear from the Fig.1 that the strongest weights for these collective modes appear for
$ \tilde k \geq 0.02 $, it implies that the momentum transfer $k$ in a light scattering 
experiments should be $ k \geq 0.18 a_z^{-1} $ for $ N = 40$
in order to pick up the strong spectral weight from the low-energy collective modes.
We also notice that the number of modes that contribute to the $S(k,\omega)$ increases with 
the chemical potential.

\begin{figure}[h]
\epsfxsize 9cm
\centerline {\epsfbox{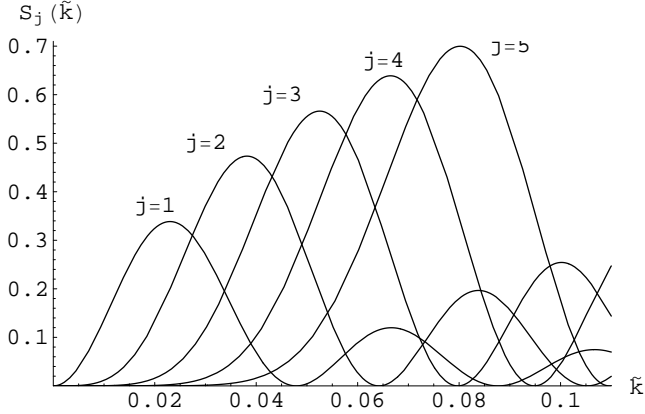}}
\vspace{0.2 cm}
\caption{Plots of $S_j( \tilde k) $ as a function of the dimensionless wave
vector $\tilde k $ for $N=40$.}
\end{figure}

For $ \tilde k \rightarrow 0$, the leading term arises from the Kohn
mode $ j =1 $, with $S_1 \sim \tilde k^2$; the next contributions arise from the terms
with $j = 2, 3 $. 
In Fig.2 we plot the dynamic structure factor $S( \tilde k,\omega)$
for the momentum transfer corresponds to $ k = 0.1 k_f $.
For finite-energy resolution we have replaced the delta function by the Lorentzian 
with a width of $ \Gamma = 0.12 \omega_z$.
\begin{figure}[h]
\epsfxsize 9cm
\centerline {\epsfbox{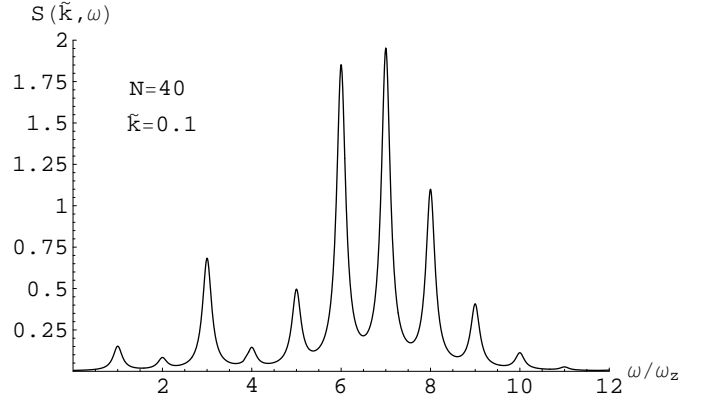}}
\vspace{0.2 cm}
\caption{Plot of the dynamic structure factor $ S(k,\omega)$ vs. $\omega/\omega_z$ at $T=0$.} 
\end{figure}

As one can see from the Fig.2, the dynamic structure factors 
has multiple peaks.
This multiple peaks are due to the underlying discrete spectrum.
Note that the energy transfer $ \omega $ is much smaller than the gap between the 
chemical potential $\mu$ and the transverse excited state ($\sim \omega_0$) and 
hence the $\omega $ can not excite the radial modes and its contribution to the dynamic structure factor
is zero. 
This behavior of the multiple peaks can be resolved in two-photon Bragg spectroscopy, as
shown by Steinhauer {\em et al.} \cite{stein}. 

{\em Bragg Spectroscopy}:
The observable in the Bragg scattering experiments is the momentum transferred to the
system. When the system is subjected to a time-dependent Bragg pulse, the additional
interaction term appears in the Hamiltonian which is given by \cite{blak},
\be
H_I(t) = \int dz \hat \psi ^{\dag} (z,t) [ V_B(t) cos(kz-\omega t)] \hat \psi (z,t).
\ee
Here, we supposed that the Bragg pulse is switched on at time $t=0$ and $k$ is along the 
$z$-direction. 

The momentum transfer from the optical potential is obtained  in Ref. \cite{tk} is given by 
\bearr
P_z(t) & = & \sum_{j,k} \hbar k < \hat \alpha_j^{\dag} (t) \hat \alpha_j (t) >
 =  \l (\frac{V_B(t)}{2 \hbar} \r )^2 \sum_j \hbar k S_j (\tilde k) \nonumber \\ 
& \times & F_j [(\omega_j - \omega),t] - F_j [(\omega_j + \omega),t],
\eearr
where $ F_j [(\omega_j \pm \omega),t] = (\frac{sin[(\omega_j \pm \omega)t/2]}{(\omega_j \pm \omega)/2})^2 $ and
$S_j (\tilde k) $ is given in Eq. (\ref{sjq}). For large $t$, $S(k,\omega) \sim P_z(t)$.
We plot $P_z(t)$ for various time in Fig.3 and shows that the multipeak spectrum in $S(k, \omega)$ can be
resolved only when the duration of the Bragg pulse is $ t >> 2\pi/\omega_z $. 
When $ t < 2\pi/\omega_z $, $P_z(t)$ reflects the dynamic structure factor calculated from
the local density approximation.

\begin{figure}[h]
\epsfxsize 9cm
\centerline {\epsfbox{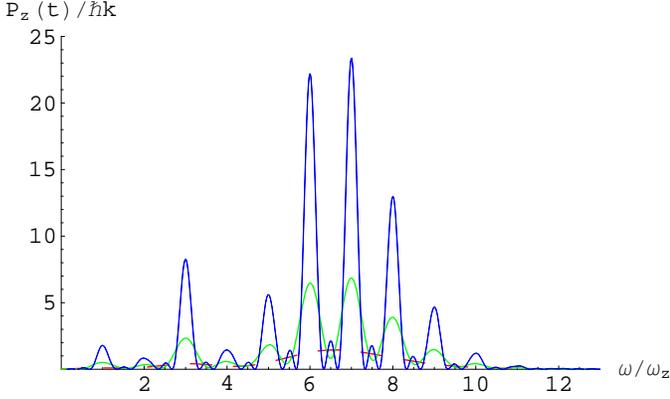}}
\vspace{0.2 cm}
\caption{(Color online). Plots of the momentum transferred $ P_z(t)$ vs. $ \omega $ at $T=0$ when
momentum transfer is $ k = 0.1 k_f$ for various time $ \omega_z t = 4.0 $ (dashed), $ \omega_z t = 10.0 $ 
(dotted) and $ \omega_z t = 19.0 $ (solid). Also, we have assumed $ V_B = 0.3 \hbar \omega_z$}.
\end{figure}

{\em Static Structure Factor}: For completeness, we also calculate
the static structure factor which can be obtained from the Fourier transform of the equal-time
density-density correlation function
\bearr
S(k) & = & \int dz e^{ik(z-z^{\prime})} \la \delta \hat n(z) \delta \hat n(z^{\prime}) \ra \nonumber \\
& = & \sum_j S_j(\tilde k) coth(\frac{\beta \hbar \omega_j}{2})
\eearr
with the long-wavelength limit $ S(\tilde k) \sim \tilde k^2 coth(\frac{\beta \hbar \omega_z}{2}) $.

\section{correlation functions and momentum distribution}
The equal-time single-particle density matrix is
defined as
\be \label{pd}
D_1(z,z^{\prime}) = \la \hat \psi^{\dag}(z,t) \hat \psi(z^{\prime},t) \ra.
\ee
The single-particle density matrix by taking care of the 
density and the phase fluctuations upto quartic term is given by
$$
D_1(z,z^{\prime}) = \sqrt{n_0(z) n_0(z^{\prime})} e^{-\frac{1}{2} F_2[z,z^{\prime}] 
+ \frac{1}{24} F_4[z,z^{\prime}] - \frac{F_d[z,z^{\prime}]}{8}},
$$
where 
\be \label{f2}
F_2[z,z^{\prime}] = \sum_{j=0}^{j_{\rm max}} \frac{1}{j}
\l [T_j(\tilde z)-T_j(\tilde z^{\prime}) \r ]^2 coth \l [\frac{\beta \hbar \omega_j}{2} \r],
\ee
and $ F_4[z,z^{\prime}] =  F_4^{\prime}[z,z^{\prime}] - 3 (F_2[z,z^{\prime}])^2 $,
where
$$
F_4^{\prime}[z,z^{\prime}]  = 
\sum_{j=0}^{j_{\rm max}} \frac{3}{j^2} \l [T_j(\tilde z)-T_j(\tilde z^{\prime}) \r ]^4 
(1 + 2N_j + 2N_j^2).
$$

and
$$ \label{fd}
F_d[z,z^{\prime}]  =  \sum_{j=0}^{j_{\rm max}} \frac{j}{4N^2}
\l [\frac{T_j(\tilde z)}{1-\tilde z^2} - \frac{T_j(\tilde z^{\prime})}{1-\tilde z^{\prime 2}} \r ]^2 
coth \l [\frac{\beta \hbar \omega_j}{2} \r ],
$$
We calculate all the summation within the phonon regime, $ \hbar \omega_j \leq \mu $ and
the upper cut-off limit, $ j_{\rm max}$, is obtained from the relation, $\mu = \hbar \omega_{j_{\rm max}} $
i.e. $ j_{\rm max} = N$. 
The normalized one-body density matrix or the phase correlation function, $ C_1(\tilde z ) $,
is defined as 
\be
C_1(z, z^{\prime}) = \frac{\la \hat \psi^{\dag} (z) \hat \psi (z^{\prime}) \ra}{\sqrt{n_0(z) n_0(z^{\prime})}} = 
e^{-\frac{1}{2} \l [F_2 - \frac{F_4}{12} + \frac{F_d}{4} \r ]}.
\ee
It is known that the strong two-body interaction induces large phase fluctuations, and
therefore we must consider the higher order term in the phase fluctuations to provide
more accurate coherence function. In fact, we have checked that the quartic phase 
fluctuation term ($F_4$) reduces the phase coherence function 
significantly (15-20 percentage).
We plot the phase coherence $ C_1(\tilde z ) $ vs. the separation $ \tilde z $ in Fig.4.
\begin{figure}[h]
\epsfxsize 9cm
\centerline {\epsfbox{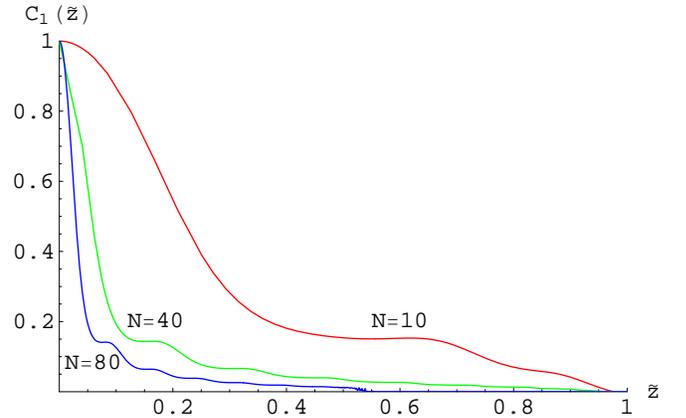}}
\vspace{0.2 cm}
\caption{(Color online). Plots of the normalized one-body density matrix $ C_1(\tilde z ) $ vs. the
the separation $\tilde z $ for various number of particles at $T=0$.}
\end{figure}

At $T=0$, the coherence function goes to zero even at small separations due to the ``fermionization" 
of the strongly interacting bosons. It has oscillatory behavior at large $z$ which is also reflected
in the momentum distributions (see Fig.5). Note that the coherence function decreases as the number
of particles increases \cite{papen}.

The momentum distribution can be obtained from Fourier transformation 
of the one-particle density matrix: 
\be
n(k) = \int_{-Z_{0}}^{Z_{0}} dz \int_{-Z_{0}}^{Z_{0}} dz^{\prime} D_1(z, z^{\prime}) e^{ik(z-z^{\prime})}.
\ee 

Fig.5 shows that the momentum distribution 
has a oscillatory behavior at large momenta which can be observed in 
the Bragg spectroscopy \cite{aspect}. Note that this hydrodynamic approximation is valid only when
$ k l_c < 1 $ $i.e.$ $ kZ_0 < 2N $. The momentum distributions for different number
of particles which is shown in Fig.5 falls within the valid hydrodynamic regime.
\begin{figure}[h]
\epsfxsize 9cm
\centerline {\epsfbox{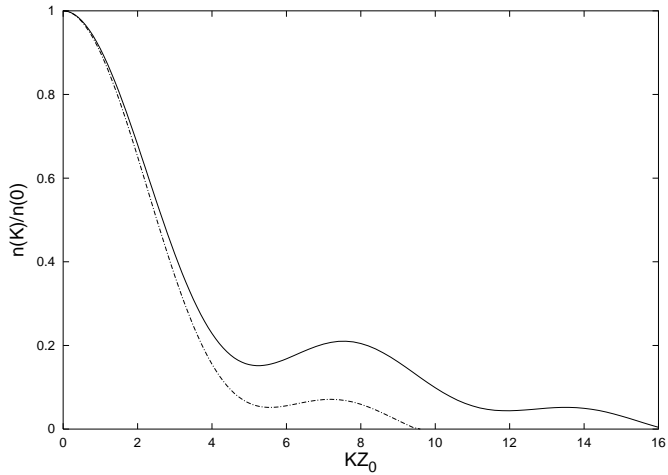}}
\vspace{0.2 cm}
\caption{Plots of the zero temperature  momentum distribution $n(k)/n(0)$ vs. momentum $ \tilde k= k Z_0$ for 
$N=10$ (solid) and $N=5$ (dot-dashed).} 
\end{figure}
The width of the momentum distributions decreases with the particle numbers \cite{girar1,papen}. 
The oscillatory behavior in the momentum distribution is also obtained by exact calculation 
in Ref. \cite{papen}.
Fig.4 shows that the single-particle correlation functions  follows the power-law decay at large 
distances:$ C_1(z) \sim 1/(\sqrt{z})$.
Also, Fig.5 shows the momentum distributions follows the power-law decay at large momentum: 
$ n(k) \sim 1/(\sqrt{k})$. These power-law decay in this system has been recently observed by 
using the exact quantum Monte Carlo techniques \cite{giorgini}.

%The two-point correlation function $C_2(z)$ reflects the nature of the statistics of the constituents
%particles in the system. 
Similar to the single particle correlation function, we also calculate
two-point correlation function at $ T=0$ by taking care of the phase fluctuations upto 
quartic term which is given by,
\be
C_2(z_1,z_2,z_3,z_4) = e^{- \frac{G_2[z_1,z_2,z_3,z_4]}{2} + \frac{G_4[z_1,z_2,z_3,z_4]}{8}},
\ee
where 
\be
G_2[z_i] = \sum_{j=1}^{N} \frac{1}{j} [ T_j(\tilde z_1) +  T_j(\tilde z_2) 
- T_j(\tilde z_3) - T_j(\tilde z_4)]^2
\ee
and 
\bearr
G_4[z_i] & = & \sum_{j=1}^{N} \frac{1}{j^2} [ T_j(\tilde z_1) +  T_j(\tilde z_2) - 
T_j(\tilde z_3) - T_j(\tilde z_4)]^4 \nonumber \\
& - & (G_2[z_i])^2.
\eearr
We plot the $C_2$ vs. the relative separation $s$ for different values of mean-distances $d$ in
Fig.6. Here, $s$ and $d$ (in units of $Z_0$) are defined as $ \tilde z_1 = (d+s)/2, 
\tilde z_2 = (- d - s)/2, \tilde z_3 = (- d + s)/2,$ and $ \tilde z_4 = ( d - s)/2$.

\begin{figure}[h]
\epsfxsize 9cm
\centerline {\epsfbox{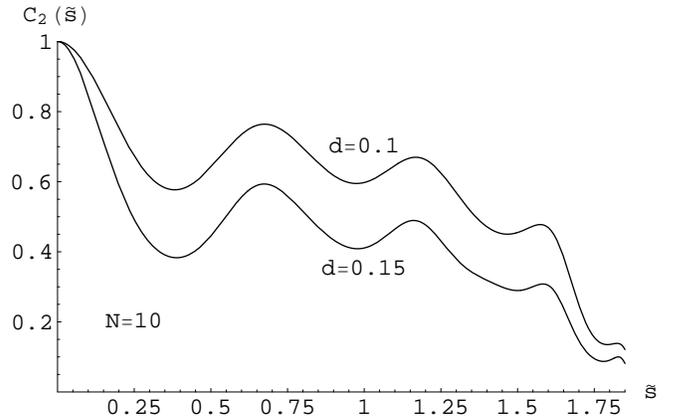}}
\vspace{0.2 cm}
\caption{Plots of the zero temperature  two-point correlation function $C_2(\tilde s)$ vs. the relative
separations $ \tilde s = s/Z_0$ for various mean-distances $d$}.
\end{figure}

The equal-times pair distribution functions shown in Fig.6 contains information on
the relative spatial distribution of pairs of hard-core bosons. We should mention that
Fig.6 is valid only when $ s > 1/(2N)$ due the hydrodynamic approximation. Actually, $C_2$
should vanish with the relative distance $s$ due to the hard-core nature of the particles.
This hydrodynamic approximation is failed to describe this short-range correlations due
to the strong interactions when $ s < 1/(2N)$. The effect of Pauli exclusion principle in real 
space is also observable as a deep valley at various relative distances $s$. It implies
that the strongly interacting bosons in 1D trap behaves as a fermions. This effect is absent
in a quasi-1D Bose gas in the mean-field regime \cite{tk2}. At large distances
the pair distribution function goes rapidly to zero and it decouples into the product 
of two-particle density profiles. Recently, the two-point correlation function has been measured
experimentally in an elongated degenerate Bose gases \cite{santos} and this method can be used
to observe the valleys in the two-point correlation function of a trapped hard-core bosons.   

\section{summary}
In this work, we have carried out a detail study of hard-core bosons in a harmonic trap
by using the hydrodynamic approximation at zero temperature. Particularly, we have calculated
dynamic structure factor, single-particle correlation
function, two-particle correlation functions and momentum distributions. We found many interesting
behavior in the observables which are absent in the trapped bosons in the mean-field
regime. All the results are obtained analytically. These results can be used to identify the 
hard-core bosons in a trap from the bosons in the mean-field regime, and also important with 
the progress of the experimental technique to achieve this strongly-interacting regime. 

I would like to thank M. A. Cazalilla for carefully reading the manuscript and valuable comments.

\end{multicols}
\end{document}